\documentclass[a4paper,11pt]{article}

\usepackage{latexsym}
\usepackage{amssymb}
\usepackage{amsfonts}
\usepackage{amsmath}

\newcommand{\bol}[1]{\boldsymbol {#1}}
\newcommand{\sss}[1]{\scriptscriptstyle {#1}}
\newcommand{\va}{\varphi}
 
\def\d{\partial}

\begin{document}

\author{Nicola Grillo\thanks{\texttt{grillo@physik.unizh.ch}} \\  \\
{\textit{Institut f\"ur Theoretische Physik, Universit\"at Z\"urich}} \\
{\textit{Winterthurerstrasse 190, CH-8057 Z\"urich, Switzerland} }}

\title{Finite One-Loop Corrections and Perturbative Gauge Invariance in Quantum Gravity 
Coupled to Photon Fields}

\maketitle

\begin{abstract}
\noindent
One-loop calculations in quantum gravity coupled to $U(1)$-Abelian fields
(photon fields) are ultraviolet finite and cutoff-free in the framework of causal perturbation theory.
We compute the photon loop graviton self-energy  and the photon self-energy in second order 
perturbation theory. The condition of perturbative gauge invariance to second order is shown and generates the 
gravitational Slavnov--Ward identities. Quantum corrections to the Newtonian potential through 
the photon loop graviton self-energy are also derived.
\vskip 7mm
{\bf Preprint:} ZU-TH 39/1999
\end{abstract}
\newpage

\tableofcontents

\newpage

\section{Introduction}
\setcounter{equation}{0}

The standard field theoretical perturbative approach to quantum gravity (see the introduction to 
this subject in~\cite{feynman} and references therein), considered as a flat space-time relativistic 
quantum field theory of {\textit{gravitons}}, massless rank-2 tensor fields, coupled to {\textit{photons}}, massless vector fields, leads to 
non-renormalizable ultraviolet (UV) divergences. These were found by means of  dimensional regularization and 
background field method~\cite{des3},~\cite{cap2},~\cite{des6} and~\cite{des8}.

A closer investigation of loop calculations in quantum field theory shows  
that the reason for the UV divergences lies basically in the fact that one 
performs  mathematically ill-defined operations, when using Feynman rules 
for closed loop graphs, because one multiplies Feynman propagators as if 
they were ordinary functions.

We show how it is possible to overcome these discouraging outcomes by applying an improved perturbation 
scheme which has as central objects the time-ordered products and as constructing principle causality. 
The $S$-matrix is constructed 
inductively as a sum of smeared operator-valued $n$-point distributions avoiding UV divergences. 

UV finiteness of $S$-matrix elements  is then a consequence of a deeper 
mathematical understanding of  how loop graph contributions have to be 
calculated.

This idea goes back 
to St\"uckelberg, Bogoliubov and Shirkov and the program (causal perturbation theory) was carried out successfully by 
Epstein and Glaser~\cite{eg} for scalar field theories and subsequently applied to QED by Scharf~\cite{scha}, to 
non-Abelian gauge theories by D\"utsch {\it et al.}~\cite{ym} and to quantum gravity~\cite{scho1}.

In this work we carry out the analysis of the coupled quantized 
Einstein--Maxwell system, Sec.\ \ref{sec:EMax}. Two main topics will be investigated in this paper: 
the gauge structure of the theory and the UV
finiteness of loop graphs in second order perturbation theory.

The first is formulated by means of a `gauge charge' that generates infinitesimal gauge variations of the fundamental 
free quantum fields. Gauge invariance of the $S$-matrix implies then a set of identities between the C-number part of 
the $2$-point distributions, Sec.~\ref{sec:CG}, which implies  the gravitational Slavnov--Ward identities 
(SWI)~\cite{capSW}.

The second is obtained by applying the `causal perturbation' scheme, 
Sec.~\ref{sec:causal}, to the calculations of the photon loop graviton 
self-energy, Sec.~\ref{sec:gravSE},  and of the photon self-energy, 
Sec.~\ref{sec:photSE}, in second order perturbation theory.  
In both cases the results are UV  finite and cutoff-free  and the inherent 
ambiguity  in the normalization of the $2$-point distribution is settled
by requiring appropriate normalization conditions for mass and coupling 
constant.

In addition, the  causal scheme preserves the gauge 
symmetries of the theory. The trace of the graviton self-energy tensor 
vanishes. This property  was  manifestly broken using dimensional 
regularization in order to extract finite results from UV divergent  
quantities~\cite{cap1}.

From the photon loop  graviton self-energy we sketch the derivation of 
quantum corrections to the Newtonian potential between two spinless massive 
bodies in the static non-relativistic limit, Sec.~\ref{sec:new}.

Gauge structure and one-loop calculations with gravitational self-coupling are not considered here, 
see~\cite{scho1},~\cite{gri3},~\cite{gri4}, and~\cite{well1}. These references contain also the notations and the 
conventions used here.

We use the unit convention: $\hbar=c=1$, Greek indices $\alpha,\beta,\ldots$ run from $0$ to $3$, whereas Latin 
indices $i,j,\ldots$ run from $1$ to $3$.

\section{Quantized Einstein--Maxwell System and Perturbative Gauge Invariance}\label{sec:EMgauge}
\setcounter{equation}{0}

\subsection{Inductive \boldmath{$S$}-Matrix Construction}\label{sec:quant}

In causal perturbation theory~\cite{scha},~\cite{aste}, one makes an ansatz for the $S$-matrix as a formal power 
series in the coupling constant, namely as a sum of smeared operator-valued distributions:
\begin{equation}
S(g)={\mathbf 1}+\sum_{n=1}^{\infty}\frac{1}{n!} \int\!\! d^{4}x_{1}
\ldots d^{4}x_{n}\, T_{n}(x_{1},\ldots, x_{n})\,g(x_{1})\cdot\ldots\cdot g(x_{n})\,  ,
\label{eq:c1}
\end{equation}
where the  Schwartz test function $g\in\mathcal{S}(\mathbb{R}^{4})$ switches adiabatically the interaction and provides a 
natural infrared cutoff in the long range part of the interaction. The $S$-matrix maps the asymptotically incoming 
free fields on the outgoing ones 
and it is possible to express the $T_{n}$'s  by means of free fields. Interacting 
quantum fields are never used in the causal scheme.

The $n$-point distribution $T_{n}$ is a  well-defined  `renormalized' time-ordered product expressed 
in terms of Wick monomials of free fields $:\!{\mathcal O}(x_{1},\ldots,x_{n})$:
\begin{equation}
T_{n}(x_{1},\ldots, x_{n})=\sum_{\sss { \mathcal{ O}}} :\!{\mathcal O}(x_{1},\ldots,x_{n})\!:\,
t_{n}^{\sss{ \mathcal O}}(x_{1}-x_{n},\ldots,x_{n-1}-x_{n})\,.
\label{eq:c1.1}
\end{equation}
The $t_{n}$'s  are C-number distributions. $T_{n}$ is constructed inductively from the first order $T_{1}(x)$, which plays
the r\^ole of the interaction Lagrangian in terms  of free fields,  and from 
the lower orders $T_{j}$, $j=2,\ldots,n-1$ by means of Poincar\'e covariance and causality. The latter leads directly 
to UV finite and cutoff-free $T_{n}$-distributions.

\subsection{Quantized Einstein--Maxwell System}\label{sec:EMax}

In the context of linearized gravity (without graviton self-interactions~\cite{scho1},~\cite{gri4}, \cite{well1}) coupled to  photon 
field~\cite{cap2}, the interaction between the  quantized 
symmetric tensor field $h^{\alpha\beta}(x)$, the graviton, and the quantized vector field $A^{\mu}(x)$, the photon, 
is considered in the background of Minkowski space-time. 

The free graviton field satisfies the wave equation
\begin{equation}
\Box\,h^{\mu\nu}(x)=0\,,
\label{eq:c10}
\end{equation}
and is quantized according to 
\begin{equation}
\big[ h^{\mu\nu}(x),h^{\alpha\beta}(y)\big]=-i\,b^{\mu\nu\alpha\beta}\,D_{0}(x-y)\,,
\label{eq:c14}
\end{equation}
where the $b$-tensor is constructed from the Minkowski metric
\begin{equation}
b^{\mu\nu\alpha\beta}:=\frac{1}{2}\left(\eta^{\mu\alpha}\eta^{\nu\beta}+\eta^{\mu\beta}\eta^{\nu\alpha}-
\eta^{\mu\nu}\eta^{\alpha\beta}\right)\,,
\label{eq:c15}
\end{equation}
and $D_{0}(x)$ is the mass-zero causal Jordan--Pauli distribution:
\begin{equation}
\begin{split}
D_{0}(x)&=D^{\sss (+)}_{0}(x)+D^{\sss (-)}_{0}(x)=\frac{1}{2\pi}\delta(x^2)\, \mathrm{sgn}(x^{0}) \\
&=\frac{i}{(2\pi)^3}\int \!\!d^{4}p\, \delta (p^2)\,
     \mathrm{sgn}(p^{0})\, e^{-i\,p\cdot x} \, .
\label{eq:c16}
\end{split}
\end{equation}
(see~\cite{gri3} and~\cite{gri4} for the details of this procedure). The additional degrees of freedom 
present in the symmetric tensor field $h^{\mu\nu}$ (gravitons are massless spin-2 particles) could be 
eliminated by imposing appropriate gauge and trace conditions, but these are disregarded in the following
and only considered later as a characterization of the physical states~\cite{gri3}.

From the point of view of Lagrangian field theory~\cite{cap2}, the Hilbert--Einstein Lagrangian 
$\mathcal{L}_{\sss HE}=-2\,\kappa^{-2}\,\sqrt{-g}\,g^{\mu\nu}\,R_{\mu\nu}$, ($\kappa^2=32\pi G$), is written by means
of the Goldberg variable $\tilde{g}^{\mu\nu}:=\sqrt{-g}\,g^{\mu\nu}$ and expanded as
$\tilde{g}^{\mu\nu}=\eta^{\mu\nu}+\kappa\,h^{\mu\nu}$ around the flat space-time  $\eta^{\mu\nu}=\mathrm{diag}(+1,-1,-1,-1)$. 
From the zeroth order, choosing the gauge $h^{\mu\nu}_{\;,\nu}=0$, one obtains~(\ref{eq:c10}).

The free photon field fulfils also the wave equation 
\begin{equation}
\Box\,A^{\mu}(x)=0\,,
\label{eq:c13}
\end{equation}
and is quantized as
\begin{equation}
\big[ A^{\mu}(x),A^{\nu}(y)\big]=+i\, \eta^{\mu\nu}\, D_{0}(x-y)\,.
\label{eq:c14.1}
\end{equation}
Also the field strength $F^{\mu\nu}:=\d^{\mu}A^{\nu}-\d^{\nu}A^{\mu}$ will be used in the following.

\subsection{Gauge Invariance and First Order Coupling \boldmath{$T_{1}^{\sss A}(x)$}}

The $U(1)_{\sss EM}$ gauge content~\cite{scha} of the photon field is formulated by means of the gauge charge 
$Q_{\sss A}$
\begin{equation}
Q_{\sss A}:=\int\limits_{x^0 =const}\!\! d^{3}x\,A^{\nu}(x)_{,\nu}{\stackrel{\longleftrightarrow}{ \partial_{x}^{0} }
 v(x)} \, ,
\label{eq:c17}
\end{equation}
where $v(x)$ is a C-number scalar field satisfying $\Box v(x)=0$. Actually, due to the generality of the gravitational 
coupling, one should also consider the coupling between $h^{\mu\nu}$ and the quantized electromagnetic ghost $v$ and  
anti-ghost $\tilde{v}$ scalar fields, but this would  have the form 
$\kappa\, \eta_{\mu\nu}h^{\mu\nu}\,\tilde{v}\Box v$ from $\sqrt{-g}\,\tilde{v}\Box v$  and therefore can be written 
as a total divergence in the 
sense of vector analysis because of the presence of two equal derivatives on three massless fields. For this reason 
the electromagnetic ghosts will not be considered here.

On the other side, the gauge content of the graviton field (which is related to the general covariance of $g_{\mu\nu}(x)$
under coordinate transformations~\cite{wald}) is formulated by the gauge charge~\cite{scho1},~\cite{well1}
\begin{equation}
Q:=\int\limits_{x^0 =const}\!\! d^{3}x\,h^{\mu\nu}(x)_{,\nu} {\stackrel{\longleftrightarrow}{ \partial_{x}^{0} }
 u_{\mu}(x)} \, .
\label{eq:c18}
\end{equation}
In order for gauge invariance to first order in pure quantum gravity to hold (see Eq. (\ref{eq:c25}))~\cite{scho1}, 
for the construction of the physical subspace~\cite{gri3} and in order to prove unitarity of the $S$-matrix on the 
physical subspace~\cite{gri3}, we need to quantize  the ghost field $u^{\mu}(x)$, together with the anti-ghost field 
$\tilde{u}^{\nu}(x)$, as fermionic vector fields
\begin{equation}
\Box u^{\mu}(x)=0\,,\quad \Box \tilde{u}^{\nu}(x)=0\,,\quad 
\big\{u^{\mu}(x),\tilde{u}^{\nu}(y)\big\}=i\, \eta^{\mu\nu}\, D_{0}(x-y) \, ,
\label{eq:c19}
\end{equation}
whereas all other anti-commutators vanish. The gauge charges generate the following infinitesimal gauge 
variations of the fundamental quantum fields:
\begin{gather}
d_Q h^{\mu\nu}(x) : =\big[ Q, h^{\mu\nu}(x)\big]=-i\, b^{\mu\nu\rho\sigma} u^{\rho}(x)_{,\sigma}\,,\nonumber\\
d_Q u^{\alpha}(x) := \big\{ Q, u^{\alpha}(x)\big\}=0\,,\quad
d_Q \tilde{u}^{\alpha}(x) := \big\{ Q, \tilde{u}^{\alpha}(x)\big\}=i\, h^{\alpha\beta}(x)_{,\beta}\,,\nonumber\\
d_{Q_{\sss A}} A^{\alpha}(x):=\big[Q_{\sss A}, A^{\alpha}(x)\big]=i\,\d_{x}^{\alpha}v(x)\,,\quad d_{Q_{\sss A}} 
F^{\alpha\beta}(x)=0\,.
\label{eq:c20}
\end{gather}
The operator $d_Q$ obeys also the Leibniz rule
\begin{equation}
d_Q (A\,B)=(d_Q A)\, B +(-1)^{n_{\sss{G}} (A)} A \, d_Q B \, ,
\end{equation}
for arbitrary operators $A$ and  $B$, where $n_{\sss{G}} (A)$ is the 
number of ghost fields minus the number of anti-ghost fields in the 
Wick monomial $A$.

For convenience of notation, the trace of the graviton field is written as $h=h^{\gamma}_{\  \gamma}$ and all Lorentz 
indices of the fields are written as superscripts whereas the derivatives acting on the fields are written as 
subscripts. All indices occurring twice are contracted by the Minkowski metric $\eta^{\mu\nu}$. We skip the 
space-time dependence if the meaning is clear.

The first order interaction between graviton field and photon fields is dictated by gauge invariance and by the 
assumption that the graviton interacts with
the traceless and conserved electromagnetic energy-momentum tensor, see below. 

Gauge invariance of the $S$-matrix means formally
\begin{equation}
\lim_{g\to 1} d_Q S(g)=0\,,\quad\text{and}\quad \lim_{g\to 1} d_{Q_{\sss A}} S(g)=0\,.
\label{eq:c24}
\end{equation}
Since the existence of the adiabatic limit $g\to 1$ in massless theories  can be 
problematic,  we consider the condition of perturbative gauge invariance to $n$-th order 
\begin{equation}
d_Q T_{n}(x_{1},\ldots,x_{n})=d_{Q_{\sss A}} T_{n}(x_{1},\ldots,x_{n})=\text{sum of divergences}\,,
\label{eq:c25}
\end{equation}
which implies $S$-matrix gauge invariance, because divergences do not contribute in the adiabatic limit $g\to 1$ 
due to partial integration and Gauss' theorem.

We define $T_{1}^{\sss A}(x)$ as
\begin{equation}
T_{1}^{\sss A}(x):=i\,\frac{\kappa}{2}\,:\!h^{\mu\nu}(x)\,b_{\mu\nu\alpha\beta}\,T^{\alpha\beta}_{\sss EM}(x)\!:\,,
\label{eq:c21}
\end{equation}
where $ T_{\sss EM}^{\alpha\beta}(x)$  corresponds to  the electromagnetic 
energy-momentum tensor and reads
\begin{equation}
T_{\sss EM}^{\alpha\beta}(x):=-F^{\alpha\gamma}(x)\,F^{\beta}_{\ \gamma}(x)+\frac{1}{4}\eta^{\alpha\beta}\,
F^{\rho\sigma}(x)\,F_{\rho\sigma}(x)\,.
\label{eq:c22}
\end{equation}
Actually, the $b$-tensor is not essential, because $T_{\sss EM}^{\alpha\beta}$ is traceless. Its origin 
lies in the Goldberg variable expansion. $T_{1}^{\sss A}(x)$ corresponds to the term of order $\kappa$ in the 
expansion of the generally covariant electromagnetic Lagrangian density
\begin{equation}
\mathcal{L}_{\sss A}=-\frac{1}{4}\,\sqrt{-g}\, F_{\mu\nu}\,F_{\alpha\beta}\,g^{\mu\alpha}\,g^{\nu\beta}\,,
\label{eq:c11}
\end{equation}
as a function of $\tilde{g}^{\mu\nu}$ around the flat space-time.

Perturbative gauge invariance to first order is readily established. $T_{1}^{\sss A}(x)$ is $U(1)_{\sss EM}$ gauge 
invariant
\begin{equation}
d_{Q_{\sss A}} T^{\sss A}_{1}(x)=0\,,
\label{eq:c26}
\end{equation}
due to $d_{Q_{\sss A}} F^{\mu\nu}(x)=0$. The $U(1)_{\sss EM}$ gauge structure of the theory is rather trivial and will
not be considered. 

Gauge invariance with respect to the gauge charge $Q$ 
\begin{equation}
d_Q T^{\sss A}_{1}(x)=\frac{\kappa}{2}\,:\!u^{\rho}(x)_{,\sigma}\,T_{\sss EM}^{\rho\sigma}(x)\!:=\d^{x}_{\sigma}
\Big(\frac{\kappa}{2}\,:\!u^{\rho}(x)\,T^{\rho\sigma}_{\sss EM}\!:\Big)=:\d^{x}_{\sigma}\,T_{1/1}^{\sigma}(x)
\label{eq:c27}
\end{equation}
holds because of $T^{\alpha\beta}_{\sss EM}(x)_{,\beta}=0$. $T_{1/1}^{\sigma}(x)$ is the so-called 
$Q$-vertex~\cite{ym},~\cite{scho1}. 
It allows us to formulate the condition of perturbative gauge invariance to the $n$-th order in a precise way :
\begin{equation}
d_{Q}T_{n}(x_{1},\ldots ,x_{n})=\sum_{l=1}^{n}\frac{\partial}{\partial x^{\nu}_{l}}\, 
T_{n/l}^{\nu}(x_{1},\ldots , x_{l},\ldots ,x_{n}) \, ,
\label{eq:c27.1}
\end{equation}
where  $T_{n/l}^{\nu}$ is the `renormalized' time-ordered product, obtained according to the inductive causal scheme, 
with one $Q$-vertex at $x_{l}$, while all other $n-1$ vertices are ordinary $T_{1}$-vertices.

\subsection{Consequences of Perturbative Gauge Invariance to Second Order}\label{sec:CG}

Without performing any calculation, from the structure of $T_{1}^{\sss A}$ we can anticipate that the two-point 
distribution describing  loop graphs has the form
\begin{equation}
T_{2}(x,y)^{\text{loops}}=:\!h^{\alpha\beta}(x)h^{\mu\nu}(y)\!:\,t_{hh}(x,y)_{\alpha\beta\mu\nu}+
:\!F^{\alpha\beta}(x)F^{\mu\nu}(y)\!:\,t_{\sss FF}(x,y)_{\alpha\beta\mu\nu}\,.\nonumber
\end{equation}
Here, the first term represents the photon loop graviton self-energy and the second term the photon self-energy.
The C-number distributions $t_{hh}$ and $t_{\sss FF}$ will be explicitly calculated in Sec.~\ref{sec:gravSE} and 
in Sec.~\ref{sec:photSE}, respectively.

Perturbative gauge invariance to second order, Eq.~(\ref{eq:c27.1}) with $n=2$, enables us to derive a set 
of identities for these distributions by comparing the distributions attached to the same external 
operators on both sides of Eq.~(\ref{eq:c27.1}), as in~\cite{du4}. Calculating $d_{Q}T_{2}(x,y)^{\text{loops}}$ 
we obtain
\begin{equation}
\begin{split}
d_{Q}T_{2}(x,y)^{\text{loops}}=&+:\!u^{\rho}(x)_{,\sigma}h^{\mu\nu}(y)\!:\,\big(-i\,b^{\alpha\beta\rho\sigma}\,
t_{hh}(x,y)_{\alpha\beta\mu\nu}\big)+\\
&+:\!h^{\alpha\beta}(x)u^{\rho}(y)_{,\sigma}\!:\,\big(-i\,b^{\mu\nu\rho\sigma}\,t_{hh}(x,y)_{\alpha\beta\mu\nu}\big)\,.
\label{eq:c28.1}
\end{split}
\end{equation}
On the other side, the loop contributions coming from $T_{2/1}^{\sigma}(x,y)$ can only be of the form
\begin{equation}
T_{2/1}^{\sigma}(x,y)=:\!u^{\rho}(x)h^{\mu\nu}(y)\!:\,t_{uh}^{\sigma}(x,y)_{\rho\mu\nu}+
:\!u^{\sigma}(x)h^{\mu\nu}(y)\!:\,t_{uh}(x,y)_{\mu\nu}\,.
\label{eq:c28.2}
\end{equation}
Applying $\d_{\sigma}^{x}$ to the expression above we find
\begin{equation}
\begin{split}
\d_{\sigma}^{x}T_{2/1}^{\sigma}(x,y)=&+:\!u^{\rho}(x)_{,\sigma}h^{\mu\nu}(y)\!:\,
\big\{t_{uh}^{\sigma}(x,y)_{\rho\mu\nu}+\eta_{\rho}^{\; \sigma}\,t_{uh}(x,y)_{\mu\nu}\big\}+\\
&+:\!u^{\rho}(x)h^{\mu\nu}(y)\!:\,\d_{\sigma}^{x}\,
\big\{ t_{uh}^{\sigma}(x,y)_{\rho\mu\nu}+\eta_{\rho}^{\; \sigma}\,
t_{uh}(x,y)_{\mu\nu}\big\}\,.
\label{eq:c28.3}
\end{split}
\end{equation}
We compare the C-number distributions in~(\ref{eq:c28.1}) and in~(\ref{eq:c28.3}) attached to the 
external operators
\begin{equation}
:\!u^{\rho}(x)_{,\sigma}h^{\mu\nu}(y)\!:\quad\text{and}\quad :\!u^{\rho}(x)h^{\mu\nu}(y)\!:\,.
\label{eq:c28.4}
\end{equation}
No such terms come from $T_{2/2}^{\sigma}(x,y)$. Therefore, we obtain the identities
\begin{equation}
\begin{split}
-i\,b^{\rho\sigma\alpha\beta}\,t_{hh}(x,y)_{\alpha\beta\mu\nu}&=\big\{t_{uh}^{\sigma}(x,y)^{\rho}_{\ \mu\nu}
+\eta^{\rho\sigma}\,t_{uh}(x,y)_{\mu\nu}\big\}\,,\\
0&=\d_{\sigma}^{x}\,\big\{t_{uh}^{\sigma}(x,y)^{\rho}_{\ \mu\nu} +\eta^{\rho\sigma}\,t_{uh}(x,y)_{\mu\nu}\big\}\,.
\label{eq:c28.5}
\end{split}
\end{equation}
By applying $\d_{\sigma}^{x}$ to the first identity and inserting the second one, we obtain
\begin{equation}
b^{\alpha\beta\rho\sigma}\,\d_{\sigma}^{x}\,t_{hh}(x,y)_{\alpha\beta\mu\nu}=0
\label{eq:c28.6}
\end{equation}
This identity has been explicitly checked in Eq.~(\ref{eq:c49}) and implies the gravitational Slavnov--Ward 
identities~\cite{cap2},~\cite{capSW} for the two-point connected Green function (see Sec.~\ref{sec:gSW})

\section{Causal Construction of the Two-Point Distribution}\label{sec:causal}
\setcounter{equation}{0}

We outline briefly the main steps in the inductive causal construction of $T_{2}(x,y)$ from the first order interaction
$T_{1}^{\sss A}(x)$. Following the inductive scheme~\cite{scha}, we first calculate
\begin{equation}
R'_{2}(x,y):=-T_{1}^{\sss A}(y)\,T_{1}^{\sss A}(x)\,,\quad A'_{2}(x,y):=-T_{1}^{\sss A}(x)\,T_{1}^{\sss A}(y)\,.
\label{eq:c2}
\end{equation}
From these two quantities, we form the causal distribution
\begin{equation}
D_{2}(x,y):=R'_{2}(x,y)-A'_{2}(x,y)=\big[T_{1}^{\sss A}(x),T_{1}^{\sss A}(y)\big]\,.
\label{eq:c3}
\end{equation}
In order to obtain $D_{2}(x,y)$, one has to carry out all possible contractions between the normally ordered $T_{1}$ 
using Wick's lemma, so that $D_{2}(x,y)$ has the following structure (see~(\ref{eq:c1.1}))
\begin{equation}
D_{2}(x,y)=\sum_{\sss { \mathcal{ O}}} :\!{\mathcal O}(x,y)\!:\,d_{2}^{\sss{ \mathcal O}}(x-y)\,.
\label{eq:c4}
\end{equation}
$d_{2}^{\sss {\mathcal O}}(x-y) $ is a numerical distributions that depends only on the relative coordinate $x-y$,
because of translation invariance. 

$D_{2}(x,y)$ contains tree (one contraction), loop (two contractions) and vacuum graph (three contractions)
contributions. Due to the presence of normal ordering, tadpole diagrams do not appear. $D_{2}(x,y)$ is causal,
\emph{i.e.} $\hbox{supp}(d_{2}^{\sss{\mathcal O}} (z))\subseteq \overline{ V^{\sss +}(z)}\cup \overline{ V^{\sss -}(z)}$, 
with $z:=x-y$.

In order to obtain $T_{2}(x,y)$ we have to split $D_{2}(x,y)$ into a retarded part, $R_{2}(x,y)$, and
an advanced part, $A_{2}(x,y)$, with respect to the coincident point $z=0$, so that 
$\hbox{supp}(R_{2}(z))\subseteq \overline{ V^{\sss +}(z)}$ and 
$\hbox{supp}(A_{2}(z))\subseteq \overline{ V^{\sss -}(z)}$.
This splitting, or `time-ordering', has to be carried out in the distributional sense with great care so that
the retarded and advanced part are well-defined and UV finite~\cite{eg},~\cite{scha}.

The splitting affects only the numerical distribution $d_{2}^{\sss {\mathcal O}}(x-y)$ and must be accomplished
according to the correct `singular order' $\omega(d_{2}^{\sss {\mathcal O}})$ which describes roughly speaking the 
behaviour of  $d_{2}^{\sss{\mathcal O}}(x-y)$ near $x-y=0$, or that of $\hat{d}_{2}^{\sss {\mathcal O}}(p)$ in the limit 
$p\to\infty$, respectively. If $\omega < 0$, then the splitting is trivial and agrees with the standard time-ordering. 
If $\omega\ge 0$,
then the splitting is non-trivial and non-unique: 
\begin{equation}
d_{2}^{\sss{\mathcal O}}(x-y)\longrightarrow r_{2}^{\sss {\mathcal O}}(x-y)+
\sum_{|a|=0}^{\omega(d_{2}^{\sss {\mathcal O}})}\,C_{a,\sss{\mathcal O}}\, D^{a}\, \delta^{\sss (4)}(x-y)\,,
\label{eq:c5}
\end{equation}
and a retarded part $r_{2}^{\sss {\mathcal O}}(x-y)$ is obtained in momentum space by means of  a
dispersion-like integral of the type
\begin{equation}
\hat{r}_{2}^{\sss {\mathcal O}}(p)=\frac{i}{2\pi}\,\int_{-\infty}^{+\infty}
\!\! dt \,\frac{\hat{d}_{2}^{\sss {\mathcal O}}(tp)}{(t-i0)^{\omega +1}\,
(1-t+i0)}\,,\quad p\in\overline{V^{\sss +}}\,,
\label{eq:cx}
\end{equation}
which requires  modifications in the case of massless theories~\cite{ym}.

Eq.~(\ref{eq:c5}) contains a local normalization ambiguity: the $C_{a,\sss{\mathcal O}}$'s are undetermined finite 
normalization constants, which multiply terms with point support $D^{a}\delta^{\sss (4)}(x-y)$ ($D^{a}$ is a partial 
differential operator). This freedom in the normalization has to be restricted by  physical conditions. For example, 
Lorentz covariance and gauge invariance will be used in our discussions.

Finally, $T_{2}(x,y)$ is obtained  by subtracting $R'_{2}(x,y)$ from $R_{2}(x,y)$ and can be written as
\begin{equation}
\begin{split}
T_{2}(x,y)+N_{2}(x,y)&=\sum_{\sss {\mathcal O}} :\!{\mathcal O}(x,y)\!: \,\bigg\{ t_{2}^{\sss {\mathcal O}}(x-y)
+\sum_{|a|=0}^{\omega(d_{2}^{\sss {\mathcal O}})}\,C_{a,\sss{\mathcal O}}\, D^{a}\,
\delta^{\sss (4)}(x-y)\bigg\} \,,
\label{eq:c7}
\end{split}
\end{equation}
where $N_{2}(x,y)$ represents the normalization terms given by the second term on the right side.

\section{Photon One-Loop  Contribution to the Graviton Self-Energy}\label{sec:gravSE}
\setcounter{equation}{0}

\subsection{Causal \boldmath$D_{2}(x,y)$-Distribution for the Photon Loop}

We now apply the causal scheme described in Sec.~\ref{sec:causal} in order to calculate the $2$-point distribution
$T_{2}(x,y)^{g\sss SE}$ that describes the photon loop graviton self-energy.

First of all, from the commutation rules~(\ref{eq:c14}) and~(\ref{eq:c14.1}) we derive the contractions between two field 
operators
\begin{gather}
C\big\{ h^{\alpha\beta}(x)\ h^{\mu\nu}(y) \big\}:=\big[h^{\alpha\beta}(x)^{\sss (-)},h^{\mu\nu}(y)^{\sss (+)}\big]=-i\,
b^{\alpha\beta\mu\nu}\, D_{0}^{\sss (+)}(x-y)\,,\nonumber \\
C\big\{ A^{\mu}(x)\ A^{\nu}(y) \big\}:=\big[A^{\mu}(x)^{\sss (-)},A^{\nu}(y)^{\sss (+)}\big]=+i\,\eta^{\mu\nu}\, 
D_{0}^{\sss (+)}(x-y)\,,
\label{eq:c29}
\end{gather}
where $(\pm)$ refers to the positive/negative frequency part of the corresponding quantity.

The distributions $R'_{2}(x,y)^{g\sss SE}$ and $A'_{2}(x,y)^{g\sss SE}$, defined in Eq.~(\ref{eq:c2}), are obtained by 
performing two photon contractions between the two first order interactions $T_{1}^{\sss A}(x)$ and $T_{1}^{\sss A}(y)$. 
After this operation, we obtain
\begin{equation}
\begin{split}
A'_{2}(x,y)^{g\sss SE}&=+:\!h^{\alpha\beta}(x)h^{\mu\nu}(y)\!:\,a'_{2}(x-y)_{\alpha\beta\mu\nu}^{g\sss SE}\,,\\
R'_{2}(x,y)^{g\sss SE}&=+:\!h^{\alpha\beta}(x)h^{\mu\nu}(y)\!:\,r'_{2}(x-y)_{\alpha\beta\mu\nu}^{g\sss SE}\,.
\label{eq:c30}
\end{split}
\end{equation}
The C-number tensorial distributions $a'_{2}(x-y)_{\alpha\beta\mu\nu}^{g\sss SE}$ and 
$r'_{2}(x-y)_{\alpha\beta\mu\nu}^{g\sss SE}$
are given by linear combinations of products between  positive/negative frequency parts of the Jordan--Pauli 
distributions~(\ref{eq:c16}) and carry four derivatives (each photon field in $T_{1}^{\sss A}$ carries one derivative). 
Therefore, the basic distribution appearing here is
\begin{equation}
D^{\sss (\pm)}_{\alpha\beta|\mu\nu}(x-y):=\d_{\alpha}^{x}\d_{\beta}^{x}D_{0}^{\sss (+)}(x-y)\cdot
   \d_{\mu}^{x}\d_{\nu}^{x}D_{0}^{\sss (+)}(x-y)\,.
\label{eq:c31}
\end{equation}
Summing up all the contributions coming from $-T_{1}^{\sss A}(x)T_{1}^{\sss A}(y)$ after performing two photon 
contractions, we have
\begin{equation}
\begin{split}
a'_{2}(x-y)_{\alpha\beta\mu\nu}^{g\sss SE}=\frac{\kappa^2}{4}\,\Big\{ &-2\eta_{\mu\nu}D^{\sss (+)}_{\alpha\rho|\beta\rho}
-2\eta_{\alpha\beta}D^{\sss (+)}_{\mu\rho|\nu\rho}-4 D^{\sss (+)}_{\alpha\beta|\mu\nu}
+4\eta_{\alpha\mu} D^{\sss (+)}_{\beta\rho|\nu\rho}+\\&+2\eta_{\alpha\nu} D^{\sss (+)}_{\beta\rho|\mu\rho}
+2\eta_{\beta\mu} D^{\sss (+)}_{\alpha\rho|\mu\rho}+\\&-\big(\eta_{\alpha\mu}\eta_{\beta\nu}+
\eta_{\alpha\nu}\eta_{\beta\mu}-
\eta_{\alpha\beta}\eta_{\mu\nu}\big)D^{\sss (+)}_{\gamma\rho|\gamma\rho}\Big\}\,,
\label{eq:c32}
\end{split}
\end{equation}
as a function of the $D_{\ldots|\ldots}^{\sss (+)}$-distributions.
The distribution $r'_{2}(x-y)_{\alpha\beta\mu\nu}^{g\sss SE}$ has the same form as above, but with 
$D_{\ldots |\ldots}^{\sss (-)}$ instead of $D_{\ldots|\ldots}^{\sss (+)}$ (because of the relation 
$D_{0}^{\sss (+)}(y-x)=-D_{0}^{\sss (-)}(x-y))$.
The $D_{\ldots|\ldots}^{\sss (\pm)}$-distribution are easily evaluated in momentum space by means of Eq.~(\ref{eq:c16}):
\begin{multline}
\hat{D}_{\alpha\beta|\mu\nu}^{\sss (\pm)}(p)=\frac{-1}{(2\pi)^4}\int\!\! d^{4}k \,\delta\big( (p-k)^2 \big)\,
\Theta\big(\pm (p^{0}-k^{0})\big)\,\delta(k^{2})\,\Theta(\pm k^{0}) \\
\times\Big[+p_{\alpha}p_{\beta}k_{\mu}k_{\nu}-p_{\alpha}k_{\beta}k_{\mu}k_{\nu}-p_{\beta}k_{\alpha}k_{\mu}k_{\nu}
+k_{\alpha}k_{\beta}k_{\mu}k_{\nu}\Big]\,,
\label{eq:c33}
\end{multline}
(see App.~1 of~\cite{gri4}), because products of Jordan--Pauli distributions 
go over into convolutions of the corresponding  Fourier transforms. Therefore, we see that we have to deal with 
integrals of the type
\begin{equation}
\begin{split}
I^{\sss (\pm)}(p)_{- / \alpha/\alpha\beta/\alpha\beta\mu/\alpha\beta\mu\nu}:=\int\!\! d^{4}k
\,&\delta\big( (p-k)^2 \big)\,
\Theta\big(\pm (p^{0}-k^{0})\big)\,\delta(k^{2})\,\Theta(\pm k^{0}) \\
&\times\big[1,k_{\alpha},k_{\alpha}k_{\beta},
k_{\alpha}k_{\beta}k_{\mu},k_{\alpha}k_{\beta}k_{\mu}k_{\nu}\big]\,.
\label{eq:c34}
\end{split}
\end{equation}
These integrals have been calculated in detail in App.~2 of~\cite{gri4} and partially in~\cite{ym}. We give here only 
the final results:
\begin{equation}
\begin{split}
I^{\sss(\pm)}(p)&=\frac{\pi}{2}\,\Theta(p^2)\,\Theta(\pm p^{0})\,,\\
I^{\sss(\pm)}(p)_{\alpha\beta}&=\frac{\pi}{6}\,\Big(p_{\alpha}p_{\beta}-\frac{p^2}{4}\,\eta_{\alpha\beta}\Big)\,
\Theta(p^2)\,
     \Theta(\pm p^{0})\,,\\
I^{\sss(\pm)}(p)_{\alpha\beta\mu}&=\frac{\pi}{8}\,\bigg(p_{\alpha}p_{\beta}p_{\mu}-\frac{p^2}{6}\Big(p_{\alpha}\,
     \eta_{\beta\mu}+p_{\beta}\,\eta_{\alpha\mu}+p_{\mu}\,\eta_{\alpha\beta}\Big)\bigg)\,
\Theta(p^2)\,\Theta(\pm p^{0})\,,\\
I^{\sss(\pm)}(p)_{\alpha\beta\mu\nu}&=\frac{\pi}{10}\,\bigg( p_{\alpha}p_{\beta}p_{\mu}p_{\nu}-\frac{p^2}{8}\,
   \Big(+p_{\alpha}p_{\beta}\,\eta_{\mu\nu}+p_{\alpha}p_{\mu}\,\eta_{\beta\nu}+p_{\alpha}p_{\nu}\,\eta_{\beta\mu}+\\
   & \qquad  +p_{\beta}p_{\mu}\,\eta_{\alpha\nu}+p_{\beta}p_{\nu}\,
    \eta_{\alpha\mu}+p_{\mu}p_{\nu}\,\eta_{\alpha\beta}\Big)+\\
&\qquad+\frac{p^4}{48}\,\Big(+\eta_{\alpha\mu}\eta_{\beta\nu}+
\eta_{\alpha\nu}\eta_{\beta\nu}+\eta_{\alpha\beta}\eta_{\mu\nu}
\Big)\bigg)\,\Theta(p^2)\,\Theta(\pm p^{0}) \,.
\raisetag{5cm}\label{eq:c36}
\end{split}
\end{equation}
From~(\ref{eq:c33}),~(\ref{eq:c34}) and~(\ref{eq:c36}), we obtain the relations for the 
$\hat{D}^{\sss (\pm)}_{\dots|\dots}$-distributions 
\begin{equation}
\begin{split}
\hat{D}^{\sss (\pm)}_{\alpha\rho|\beta\rho}(p)&=\frac{-1}{(2\pi)^4}\,\big[p_{\alpha}p_{\rho}\,
I^{\sss (\pm)}(p)_{\rho\beta}
-p_{\rho}\, I^{\sss (\pm)}(p)_{\rho\alpha\beta}\big]\\
&=\frac{-\pi}{48(2\pi)^4}\,\big[2 p^2 p_{\alpha} p_{\beta} +p^4 \eta_{\alpha\beta}\big]\,\Theta(p^2)\,
\Theta(\pm p^0)\,,\\
\hat{D}^{\sss (\pm)}_{\gamma\rho|\gamma\rho}(p)&=\frac{-1}{(2\pi)^4}\,\big[p_{\gamma}p_{\rho}\,
I^{\sss (\pm)}(p)_{\gamma\rho}\big]\\
&=\frac{-\pi}{8(2\pi)^4}\,p^4 \,\Theta(p^2)\,\Theta(\pm p^0)\,.
\label{eq:c36.1}
\end{split}
\end{equation}
Therefore, from~(\ref{eq:c30})  with $
\Theta(p^2)\,\Theta(-p^0)-\Theta(p^2)\,\Theta(+p^0)=-\Theta(p^2)\,\mathrm{sgn}(p^0)$  we find
\begin{equation}
D_{2}(x,y)^{g\sss SE}=:\!h^{\alpha\beta}(x)h^{\mu\nu}(y)\!:\,d_{2}(x-y)_{\alpha\beta\mu\nu}^{g\sss SE}\,.
\label{eq:c37}
\end{equation}
With Eqs.~(\ref{eq:c32}),~(\ref{eq:c36.1}) and after symmetrization in 
$(\alpha\beta)\leftrightarrow(\mu\nu)$ the C-number tensorial $d_{2}$-distribution reads in momentum space
\begin{equation}
\begin{split}
\hat{d}_{2}(p)^{g\sss SE}_{\alpha\beta\mu\nu}=\frac{\kappa ^2 \pi}{960 (2\pi)^4}\,\Big[
&-16\:  p_{\alpha}p_{\beta}p_{\mu}p_{\nu} 
 -8 \:  p^2\big( p_{\alpha}p_{\beta}\eta_{\mu\nu}+p_{\mu}p_{\nu}\eta_{\alpha\beta}\big)+\\
&+12\:  p^2 \big(p_{\alpha}p_{\mu}\eta_{\beta\nu}+p_{\alpha}p_{\nu}\eta_{\beta\mu}+p_{\beta}p_{\mu}\eta_{\alpha\nu}+
                   p_{\beta}p_{\nu}\eta_{\alpha\mu}\big)+ \\
&-12\:  p^4 \big(\eta_{\alpha\mu}\eta_{\beta\nu}+\eta_{\alpha\nu}\eta_{\beta\mu}\big)\\
&+8\:  p^4 \eta_{\alpha\beta}\eta_{\mu\nu}\Big]\,\Theta(p^2)\,\mathrm{sgn}(p^0)  \,.
\label{eq:c38}
\end{split}
\end{equation}
For simplicity, we use the shorthand notation
\begin{gather}
\hat{d}_{2}(p)^{g\sss SE}_{\alpha\beta\mu\nu}=\Upsilon\,\hat{P}(p)^{g\sss SE}_{\alpha\beta\mu\nu}\,\hat{d}(p)\,,
\nonumber\\
\hat{P}(p)^{g\sss SE}_{\alpha\beta\mu\nu}:=\big[-16,-8,+12,-12,+8\big]\,,
\label{eq:c39}
\end{gather}
where $\Upsilon:=\kappa^2 \pi/(960 (2\pi)^4)$, and $\hat{d}(p):=\Theta(p^2)\,\mathrm{sgn}(p^0)$ (this scalar 
distribution is typical for a loop with massless particles) and the numerical coefficients always refer to the polynomial 
structure as in Eq.~(\ref{eq:c38}). The polynomial $\hat{P}(p)^{g\sss SE}_{\alpha\beta\mu\nu}$  has degree four in $p$, 
because of the four derivatives present on the two contracted photon lines.

\subsection{Singular Order, Distribution Splitting  and Graviton Self-Energy Tensor}\label{sec:42}

According to the inductive construction of $T_{2}(x,y)$, Sec.~\ref{sec:causal}, the next step is the splitting of  
$D_{2}(x,y)$ into a retarded part $R_{2}(x,y)$ and an advanced part $A_{2}(x,y)$. In this procedure the singular  order 
of the distribution $D_{2}(x,y)$ plays an essential r\^ole. From~(\ref{eq:c38}) it follows that $\omega(\hat{d}_{2})=4$, 
because of the presence of the polynomial of degree four in $p$. 

A general formula for the singular order of any distribution in linearized gravity coupled to photon fields can be 
found by considering in the $n$-th order of perturbation theory an arbitrary $n$-point distribution
\begin{equation}
T^{\sss G}_{n}(x_{1},\ldots ,x_{n})
= :\!\prod_{j=1}^{n_h}h(x_{k_{j}})\,\prod_{i=1}^{n_{\sss A}}A(x_{m_{i}})\!:\,t^{\sss G}_{n}(x_{1},\ldots ,x_{n})\,.
\label{eq:c40}
\end{equation}
This $T_{n}^{\sss G}$ corresponds to a graph $G$ with $n_{h}$ external graviton lines and  $n_{\sss A}$ external 
photon lines. Then we state that the  singular order of $G$ is given by
\begin{equation}
\omega(G)\le 4-n_{h}-n_{\sss A}-d+n\,.
\label{eq:c41}
\end{equation}
Here, $d$ is the number of derivatives on the external field operators in~(\ref{eq:c40}). The explicit presence of the 
order of perturbation theory renders the theory `non-normalizable': the number of free  undetermined, but 
finite normalization terms in Eq.~(\ref{eq:c7}) increases with the order of 
perturbation theory, that is the theory has a weaker predictive power but 
it is still well-defined in the sense of UV finiteness. The proof of~(\ref{eq:c41}) has the same structure
as in QED, Yang--Mills theories and  pure quantum gravity, see therefore~\cite{scha},~\cite{ym} and~\cite{gri4},
respectively.

In the case of the graviton self-energy contribution, Eq.~(\ref{eq:c37}), we obtain from~(\ref{eq:c41}) 
the correct result $\omega(\hat{d}_{2})=4$, being $n_{h}=2$, $n_{\sss A}=0$, $d=0$ and $n=2$. The singular order of a 
distribution remains unchanged after distribution splitting. 

Because of the decomposition~(\ref{eq:c39}), it suffices to split $\hat{d}(p)$, which has $\omega(\hat{d})=0$, in order 
to obtain a retarded part $\hat{r}(p)$. The full retarded distribution is then given by  
$\hat{r}_{2}(p)_{\alpha\beta\mu\nu}^{g\sss SE}=\Upsilon\hat{P}(p)^{g\sss SE}_{\alpha\beta\mu\nu}\hat{r}(p)$. 
The ambiguity in the normalization  appearing in the splitting formula~(\ref{eq:c5}) will be discussed in Sec.~\ref{sec:norm1}.

The splitting of the scalar distribution $\hat{d}(p)$ was already carried out in~\cite{ym}, using a modification of 
the formula~(\ref{eq:cx}) for the retarded part, see also~\cite{gri4}. We 
quote only the result for the analytic continuation to $p\in\mathbb{R}^4$ of the retarded part:
\begin{equation}
\hat{r}(p)^{an}=\frac{i}{2\pi}\log\left( \frac{-p^2 - i\,p^{0} 0}{M^2}\right)\,,
\label{eq:c42}
\end{equation}
where $M>0$ is a scale invariance breaking mass. Then 
\begin{gather}
R_{2}(x,y)^{g\sss SE}=:\!h^{\alpha\beta}(x)h^{\mu\nu}(y)\!:\,r_{2}(x-y)_{\alpha\beta\mu\nu}^{g\sss SE}\,,\nonumber\\
\hat{r}_{2}(p)^{g\sss SE}_{\alpha\beta\mu\nu}=i\,\frac{\Upsilon}{2\pi}\,\hat{P}(p)^{g\sss SE}_{\alpha\beta\mu\nu}\,
\log\left( \frac{-p^2 - i\,p^{0} 0}{M^2}\right)\,.
\label{eq:c43}
\end{gather}
The distribution $T_{2}(x,y)$ is obtained from $R_{2}(x,y)$ by subtracting $R'_{2}(x,y)$, Eq.~(\ref{eq:c30}). 
This subtraction affects only the scalar distribution. Since $\hat{r}'(p)=-\Theta(p^2)\,\Theta(-p^0)$, we obtain
\begin{equation}
\hat{t}(p)=\hat{r}(p)^{an}-\hat{r}'(p)=\frac{i}{2\pi}\log\left( \frac{-p^2 -i\,0}{M^2}\right)\,.
\label{eq:c44}
\end{equation}
Therefore, the $2$-point distribution for the photon loop graviton self-energy reads
\begin{equation}
\begin{split}
T_{2}(x,y)^{g\sss SE}&=:\!h^{\alpha\beta}(x)h^{\mu\nu}(y)\!:\,i\,\Pi(x-y)_{\alpha\beta\mu\nu}^{g\sss SE}\,,\\
i\,\hat{\Pi}(p)_{\alpha\beta\mu\nu}&=i\,\Xi\,\big[-16,-8,+12,-12,+8\big]\,\log\left( \frac{-p^2 - i\, 0}{M^2}
\right)\,,
\label{eq:c45}
\end{split}
\end{equation}
with $\Xi:=\frac{1}{2\pi}\Upsilon$. The most important aspect of this result lies in the fact that the 
obtained $T_{2}(x,y)^{g\sss SE}$-distribution is UV finite without the introduction of 
counterterms~\cite{des3},~\cite{cap2} and~\cite{des6} and cutoff-free. Note that $M$ is a normalization
constant and not a cutoff.

\subsection{Slavnov--Ward Identities from Perturbative Gauge Invariance}\label{sec:gSW}

We investigate the gauge properties of the graviton self-energy tensor. For simplicity let us denote by
$[A,B,C,E,F]$ the five numerical coefficients of  $\Pi(x-y)_{\alpha\beta\mu\nu}$ given as in Eq.~(\ref{eq:c38}). 

First of all, $T_{2}(x,y)^{g\sss SE}$ satisfies the condition of perturbative gauge invariance to second 
order~(\ref{eq:c27.1}). Using the infinitesimal operator gauge variation~(\ref{eq:c20}), we obtain
\begin{equation}
\begin{split}
d_{Q}T_{2}(x,y)^{g\sss SE}=&+\d_{\sigma}^{x}\Big(:\!u^{\rho}(x)h^{\mu\nu}(y)\!:\,\big[+b^{\alpha\beta\rho\sigma}\,
             \Pi(x-y)_{\alpha\beta\mu\nu}\big]\Big)+\\
&+\d_{\sigma}^{y}\Big(:\!h^{\alpha\beta}(x)u^{\rho}(y)\!:\,\big[+b^{\mu\nu\rho\sigma}\,
          \Pi(x-y)_{\alpha\beta\mu\nu}\big]\Big)\,,
\label{eq:c47}
\end{split}
\end{equation}
because, after working out explicitly in momentum space, the necessary condition for~(\ref{eq:c47}) to hold,
\begin{equation}
b^{\alpha\beta\rho\sigma}\,p_{\sigma}\,\hat{\Pi}(p)_{\alpha\beta\mu\nu}=0\,,
\label{eq:c48}
\end{equation}
corresponds to the identity~(\ref{eq:c28.6}) and is satisfied by $\hat{\Pi}(p)_{\alpha\beta\mu\nu}$ of 
Eq.~(\ref{eq:c45}). In terms of the $A,\ldots, F$ coefficients,~(\ref{eq:c48}) implies
\begin{equation}
C+E=0\,,\quad A-2B=0\,,\quad B-2E-2F=0\,.
\label{eq:c49}
\end{equation}
and these relations are satisfied by the polynomial in~(\ref{eq:c45}). 

We show now that these relations implies the gravitational Slavnov--Ward identities (SWI)~\cite{cap2},~\cite{capSW}. 
We construct the $2$-point connected Green function with one photon loop
\begin{equation}
\hat{G}(p)_{\alpha\beta\mu\nu}^{\sss [2]}:=b_{\alpha\beta\gamma\delta}\,\hat{D}_{0}^{\sss F}(p)\,
\hat{\Pi}(p)^{\gamma\delta\rho\sigma}\,b_{\rho\sigma\mu\nu}\,\hat{D}_{0}^{\sss F}(p)\,,
\label{eq:c50}
\end{equation}
with the Feynman propagator $\hat{D}_{0}^{\sss F}(p)=(2\pi)^{-2}(-p^2-i0)^{-1}$. The gravitational SWI 
reads
\begin{equation}
p^{\alpha}\hat{G}(p)_{\alpha\beta\mu\nu}^{\sss [2]}=0\,,
\label{eq:c51}
\end{equation}
namely the $2$-point connected Green function is transversal. Eq.~(\ref{eq:c51}) implies the relations
\begin{equation}
C+E=0\,,\quad A-2B=0\,,\quad \frac{A}{4}+C-F=0\,.
\label{eq:c52}
\end{equation}
Analysis of~(\ref{eq:c49}) and~(\ref{eq:c52}) shows that the condition of perturbative gauge invariance is equivalent to 
the gravitational SWI. In addition, the photon loop is transversal also without the $b$-tensors that come from the 
graviton propagators:
\begin{equation}
p^{\alpha}\,\hat{\Pi}(p)_{\alpha\beta\mu\nu}=0\,,
\label{eq:c55}
\end{equation}
because the equivalent relations 
\begin{equation}
C+E=0\,,\quad B+F=0\,,\quad A+B+2C=0\,.
\label{eq:c56}
\end{equation}
are fulfilled by~(\ref{eq:c45}). This property follows directly from our perturbative invariance 
condition~(\ref{eq:c28.6}), because the trace of $\hat{\Pi}(p)_{\alpha\beta\mu\nu}$ vanishes (see below).

The last two properties of $\hat{\Pi}(p)_{\alpha\beta\mu\nu}$ concern its trace and are related to conformal 
transformations. Conformal invariance is manifest in the 
vanishing of the trace of the Maxwell energy-momentum tensor in four dimension. Basically, the photon loop 
graviton self-energy consists of a  `time-ordered' product of two such traceless photon energy-momentum 
tensors~(\ref{eq:c22}). Therefore, in addition to the identities~(\ref{eq:c48}), it is expected that
$\hat{\Pi}(p)_{\alpha\beta\mu\nu}$ and $\hat{G}(p)_{\alpha\beta\mu\nu}^{\sss [2]}$ are traceless.
The condition
\begin{equation}
\eta^{\alpha\beta}\,\hat{\Pi}(p)_{\alpha\beta\mu\nu}=0\,,
\label{eq:c57}
\end{equation}
namely that the photon loop graviton self-energy tensor is traceless, implies the relations
\begin{equation}
B+2E+4F=0\,,\quad A+4B+4C=0\,.
\label{eq:c58}
\end{equation}
These are satisfied by~(\ref{eq:c45}). 
Note that~(\ref{eq:c58}) does not follow from the perturbative gauge invariance
conditions~(\ref{eq:c49}). In addition, in our approach, also the trace of the $2$-point connected Green 
function vanishes:
\begin{equation}
\eta^{\alpha\beta}\,\hat{G}(p)_{\alpha\beta\mu\nu}^{\sss [2]}=0\,,
\label{eq:c59}
\end{equation}
because the equivalent relations
\begin{equation}
\frac{A}{2}+3B+2C+2E+4F=0\,,\quad A+4B+4C=0\,.
\label{eq:c60}
\end{equation}
are satisfied by the coefficients of the photon loop self-energy tensor in Eq.~(\ref{eq:c45}). This is in 
sharp contrast to the calculation  in~\cite{cap2},~\cite{cap1}, performed within the dimensional 
regularization scheme. This scheme generates conformal trace anomalies~\cite{cap1.1}, because the trace 
operation is not dimensional invariant. As a consequence, the extraction of the finite part from UV 
divergent quantities breaks this symmetry. 

Causal perturbation theory not only provides us with UV finite 
results but preserves the underlying symmetries of the theory such as the SWI and the vanishing of the 
trace of the self-energy tensor. Invariance under conformal transformations is broken by the presence of 
the mass scale $M$ in the logarithm of Eq.~(\ref{eq:c45}). If necessary, we could exploit the ambiguity in the  
normalization  to restore it, see Sec.~\ref{sec:norm1}.

\subsection{Freedom in the Normalization of the Graviton Self-Energy}\label{sec:norm1}

We still have to discuss the ambiguity in the splitting procedure, namely the appearance of undetermined local 
normalization terms.

Having singular order four, the $2$-point photon loop graviton self-energy contribution $T_{2}(x,y)^{g\sss SE}$ admits a 
general normalization  of the form
\begin{gather}
N_{2}(x,y)^{g\sss SE}=:\!h^{\alpha\beta}(x)h^{\mu\nu}(y)\!:\,i\,N(\d_{x},\d_{y})_{\alpha\beta\mu\nu}\,
\delta^{\sss (4)}(x-y)\,,\nonumber \\
\hat{N}(p)_{\alpha\beta\mu\nu}=\hat{N}(p)^{\sss (0)}_{\alpha\beta\mu\nu}+\hat{N}(p)^{\sss (2)}_{\alpha\beta\mu\nu}
+\hat{N}(p)^{\sss (4)}_{\alpha\beta\mu\nu}\,,
\label{eq:c62}
\end{gather}
where the odd terms are excluded by  parity. $\hat{N}(p)^{\sss (i)}_{\alpha\beta\mu\nu}$ is a polynomial in $p$ of 
degree $i$ with  $i=0,2,4$. We assume in addition that only scalar constants should be considered, because vector-valued 
or tensor-valued constants may endanger Lorentz covariance.  Taking the symmetries of $\Pi(x-y)_{\alpha\beta\mu\nu}$ 
into account, we make the following ansatz
\begin{equation}
\begin{split}
&\hat{N}(p)^{\sss (0)}_{\alpha\beta\mu\nu}=\Xi\,\big[+c_{1}\,\big(\eta_{\alpha\mu}\eta_{\beta\nu}+
        \eta_{\alpha\nu}\eta_{\beta\mu}\big)+c_{2}\,\eta_{\alpha\beta}\eta_{\mu\nu} \big]\,,\\
&\begin{split}\hat{N}(p)^{\sss (2)}_{\alpha\beta\mu\nu}=\Xi\,\big[&+c_{3}\,\big( p_{\alpha}p_{\beta}\eta_{\mu\nu}
   +p_{\mu}p_{\nu}\eta_{\alpha\beta}\big)+\\
 & +c_{4}\,\big( p_{\alpha}p_{\mu}\eta_{\beta\nu}+ p_{\alpha}p_{\nu}\eta_{\beta\mu}+p_{\beta}p_{\mu}\eta_{\alpha\nu}+
   p_{\beta}p_{\nu}\eta_{\alpha\mu}\big)+\\
 & +c_{5}\,p^{2}\,\big(\eta_{\alpha\mu}\eta_{\beta\nu}+\eta_{\alpha\nu}\eta_{\beta\mu}\big)
   +c_{6}\,p^{2}\,\eta_{\alpha\beta}\eta_{\mu\nu}\big]\,,
\end{split}
\\
&\hat{N}(p)^{\sss (4)}_{\alpha\beta\mu\nu}=\Xi\,\big[c_{7},c_{8},c_{9},c_{10},c_{11}\big]\,.
\label{eq:c63}
\end{split}
\end{equation}
$c_{1},\ldots,c_{11}$ are undetermined real numbers. The normalization polynomials have to fulfil the same symmetries as
$\hat{\Pi}_{\alpha\beta\mu\nu}$, namely:~(\ref{eq:c49}),~(\ref{eq:c56}),~(\ref{eq:c58}) and~(\ref{eq:c60}). 
These reduce 
the choice to 
\begin{gather}
\hat{N}(p)^{\sss (0)}_{\alpha\beta\mu\nu}=0\,,\qquad \hat{N}(p)^{\sss (2)}_{\alpha\beta\mu\nu}=0\,,\nonumber\\
\hat{N}(p)^{\sss (4)}_{\alpha\beta\mu\nu}=\Xi\,\frac{c_{11}}{4}\,\big[-16,-8,+12,-12,+8\big]\,.
\label{eq:c64}
\end{gather}
Since $\hat{N}(p)^{\sss (4)}_{\alpha\beta\mu\nu}$ has the same structure as $\hat{\Pi}_{\alpha\beta\mu\nu}$, we absorb 
the undetermined parameter $c_{11}$ in the mass scale $M$ appearing in Eq.~(\ref{eq:c45}) through the rescaling
$c_{11}=4\log(M^2 /M^2_{0})$. 
The whole freedom in the normalization  is reduced to the mass parameter $M_{0}$ in the
logarithm owing to the symmetries that the self-energy tensor and its normalization  have to fulfil. 

This normalization automatically preserves graviton mass-  and coupling constant-normalizations.
If one sums up the infinite series with an increasing 
number  of photon loop graviton self-energy insertions, the vanishing of 
$\hat{N}(p)^{\sss (0)}_{\alpha\beta\mu\nu}$ implies that the graviton mass is 
not shifted by quantum corrections and the vanishing of 
$\hat{N}(p)^{\sss (2)}_{\alpha\beta\mu\nu}$ implies that the coupling constant 
$\kappa$ is not  changed by quantum corrections.

As pointed out at the end of Sec.~\ref{sec:gSW}, we could exploit this  normalization  to restore 
conformal invariance. Through a rescaling of $M_{0}$ we can compensate the variation of  $p^2$ under such 
a transformation. The drawback lies in the arbitrariness  of this operation.

\subsection{Corrections to the Newtonian Potential}
\label{sec:new}

As pointed out in~\cite{dono1},~\cite{dono2},~\cite{hambe} and~\cite{sha}, massless particle loop 
corrections to  the graviton propagator leads to quantum corrections of  the Newtonian potential between 
massive spinless bodies in the static non-relativistic limit. These can be appropriately defined by 
considering the whole set of diagrams in the 
scattering $\va_{1}\va_{2}\to\va_{1}\va_{2}$, where $\va_{i}$ represents a  scalar field of mass $m_{i}$, of the  
order $\kappa^4$ (\emph{i.e.} $\sim G^2$). Then one isolates the non-local contributions. These  lead to $r^{-2}$ and 
$r^{-3}$ corrections to the Newtonian potential $V(r)=-Gm_{1}m_{2} r^{-1}$ in the static non-relativistic limit. For 
this purpose, the logarithm-dependent result~(\ref{eq:c45})  generates  the correction
\begin{equation}
V(r)=-\,G\,\frac{m_{1}m_{2}}{r}\,\left( 1+\frac{G\hbar}{c^{3}\pi}\,\frac{8}{15}\frac{1}{r^{2}}\right)\,,
\label{eq:c65}
\end{equation}
as calculated in~\cite{gri4}, where $\hbar$ and $c$ are put back in the expression and the relevant 
length scale appears to be the Planck 
length $l_{p}=\sqrt{G\hbar/c^{3}}$. 

Note that the mass scale $M$ in Eq.~(\ref{eq:c45}) is irrelevant, because it 
contributes only to local terms $\sim\delta^{\sss (3)}(\bol{x})$. Therefore,
the still remaining freedom in the normalization, namely the choice of $M$, is
irrelevant to physical quantum corrections of the Newtonian potential.

The  correction in Eq.~(\ref{eq:c65}) is 
only a partial one,  because we have taken into account only the photon loop 
graviton self-energy contribution and not the complete set of  diagrams of 
order $\kappa^4$  contributing to these corrections, as, for example, the 
vertex correction or the double scattering. Therefore we cannot make any 
statement on the absolute sign and magnitude of the numerical factor  in 
Eq.~(\ref{eq:c65}). 
To our knowledge, photon loop corrections to the exchanged graviton have not 
been considered yet.

\section{Photon Self-Energy}\label{sec:photSE}
\setcounter{equation}{0}

We now turn to the calculation of the photon self-energy. Since the main features of the causal scheme have already 
been presented, we can go on more speedily. Moreover this graph is not interesting from the point of view of
gauge invariance: it is trivially gauge invariant due to $d_{Q_{\sss A}} F^{\alpha\beta}=0$.

In order to obtain the relevant $A'_{2}(x,y)$ and $R'_{2}(x,y)$, we carry out one graviton and 
one field-strength contraction 
between $T_{1}^{\sss A}(x)$ and $T_{1}^{\sss A}(y)$ and obtain
\begin{equation}
A'_{2}(x,y)=:\!F^{\alpha\gamma}(x)F^{\mu\rho}(y)\!:\,a'_{2}(x-y)_{\alpha\gamma|\mu\rho}\,,
\label{eq:c66}
\end{equation}
with
\begin{multline}
a'_{2}(x-y)_{\alpha\gamma|\mu\rho}:=\frac{\kappa^2}{4}\,\big(-\eta_{\gamma\rho}\,D^{\sss (+)}_{\cdot|\mu\alpha}-
\eta_{\gamma\mu}\,D^{\sss (+)}_{\cdot|\alpha\rho}-\eta_{\alpha\rho}\,D^{\sss (+)}_{\cdot|\gamma\mu}+\\
-5\eta_{\alpha\mu}\,D^{\sss (+)}_{\cdot|\gamma\rho}+2\eta_{\gamma\alpha}\,D^{\sss (+)}_{\cdot|\mu\rho}+2 
 \eta_{\mu\rho}\,D^{\sss (+)}_{\cdot|\gamma\alpha}\big)(x-y)\,,
\label{eq:c67}
\end{multline}
where we have as in~(\ref{eq:c33})
\begin{equation}
\hat{D}_{\cdot|\mu\nu}^{\sss (+)}(p)=\frac{1}{(2\pi)^4}\int\!\! d^{4}k \,\delta\big( (p-k)^2 \big)\,
\Theta(p^{0}-k^{0})\,\delta(k^{2})\,\Theta(k^{0})\, k_{\mu}k_{\nu}\,.
\label{eq:c67.1}
\end{equation}
The expression in~(\ref{eq:c67}) has to be antisymmetrized in $\alpha\leftrightarrow\gamma$ and in 
$\mu\leftrightarrow\rho$, because these  manifest antisymmetries in~(\ref{eq:c66}) get lost using the
$D_{\dots|\dots}^{\sss (\pm)}$-distributions. Therefore, with~(\ref{eq:c34}) we find in momentum space
\begin{equation}
\begin{split}
\hat{a}'_{2}(p)_{\alpha\gamma|\mu\rho}=\frac{\kappa^2 \pi}{48(2\pi)^4}\,\big[&+2p_{\alpha}p_{\rho}\eta_{\gamma\mu}
-2p_{\gamma}p_{\rho}\eta_{\alpha\mu}-2p_{\alpha}p_{\mu}\eta_{\gamma\rho}+2p_{\gamma}p_{\mu}\eta_{\alpha\rho}+\\
&-p^2 \eta_{\gamma\mu}\eta_{\alpha\rho}+p^2 \eta_{\alpha\mu}\eta_{\gamma\rho}\big]\,\Theta(p^2)\,\Theta(p^0)\,.
\label{eq:c68}
\end{split}
\end{equation}
The result for $R'_{2}(x,y)$ is analogous to that of $A'_{2}(x,y)$: the C-number tensor distribution 
$r'_{2}(x-y)_{\alpha\gamma|\mu\rho}$ depends on the $D^{\sss (-)}_{\ldots|\ldots}$-distributions so that in momentum 
space the factor $\Theta(-p^0)$ is present instead of $\Theta(p^0)$. The causal $D_{2}$-distribution is then given by
\begin{gather}
D_{2}(x,y)^{p\sss SE}=:\!F^{\alpha\gamma}(x)F^{\mu\rho}(y)\!:\,
d_{2}(x-y)^{p\sss SE}_{\alpha\gamma|\mu\rho}\,,\nonumber\\
\hat{d}(p)^{p\sss SE}_{\alpha\gamma|\mu\rho}=\frac{-\kappa^2\pi}{48(2\pi)^4}\,
\big[\text{the same as in~(\ref{eq:c68})}\big]\,
\Theta(p^2)\,\mathrm{sgn}(p^0)\,.
\label{eq:c69}
\end{gather}
From Eq.~(\ref{eq:c41}) or from direct inspection, we ascertain that the singular order of the distribution is two. 
The distribution splitting is the same as in Sec.~\ref{sec:42}, Eq.~(\ref{eq:c44}), because the loop consists of 
massless particles. Therefore, the  $T_{2}(x,y)$-distribution corresponding to the photon self-energy reads
\begin{gather}
T_{2}(x,y)^{p\sss SE}=:\!F^{\alpha\gamma}(x)F^{\mu\rho}(y)\!:\,
\big(-i\,\Pi(x-y)^{p\sss SE}_{\alpha\gamma|\mu\rho}\big)\,,\nonumber\\
\hat{\Pi}_{2}(p)^{p\sss SE}_{\alpha\gamma|\mu\rho}=\frac{\kappa^2\pi}{48(2\pi)^5}\,
\big[\text{the same as in~(\ref{eq:c68})}\big]\,
\log\left(\frac{-p^2-i0}{M^2}\right)\,.
\label{eq:c70}
\end{gather}
$U(1)_{\sss EM}$-perturbative gauge invariance to second order holds trivially:
\begin{equation}
d_{Q_{\sss A}} T_{2}(x,y)^{p\sss SE}=0\,.
\end{equation}
The  photon self-energy $2$-point distribution can be written also in terms of the photon field $A^{\mu}$: 
\begin{equation}
\begin{split}
T_{2}(x,y)^{p\sss SE}&=:\!A^{\gamma}(x)_{,\alpha}A^{\rho}(y)_{,\mu}\!:\,\big(-4\,i\,
\Pi(x-y)^{p\sss SE}_{\alpha\gamma|\mu\rho}\big)\\
&=:\!A^{\gamma}(x)A^{\rho}(y)\!:\,\big( -i\,\Pi(x-y)_{\gamma\rho}\big)\,,
\label{eq:c71}
\end{split}
\end{equation}
by disregarding non-contributing divergences (in the adiabatic limit $g\to 1$ of Eq.~(\ref{eq:c1})). 
Then $T_{2}(x,y)^{p\sss SE}$ has the same 
structure as the photon  self-energy in QED~\cite{scha} and the `reduced' photon self-energy tensor appearing 
Eq.~(\ref{eq:c71}) is
\begin{equation}
\begin{split}
\hat{\Pi}(p)_{\gamma\rho}&=4 \,p^{\mu}p^{\alpha}\,\hat{\Pi}(p)_{\alpha\gamma|\mu\rho}\\
&=\frac{\kappa^2\pi}{12(2\pi)^5}\,\big[p^2 p_{\gamma}p_{\rho}-p^4 \eta_{\gamma\rho}\big]
\log\left(\frac{-p^2-i0}{M^2}\right)
\label{eq:c72}
\end{split}
\end{equation}
and satisfies the Ward identity
\begin{equation}
p^{\gamma}\,\hat{\Pi}(p)_{\gamma\rho}=0
\label{eq:c74}
\end{equation}
as in QED.

As a last point, we discuss the normalization of $T_{2}(x,y)^{p\sss SE}$ in the second line of  Eq.~(\ref{eq:c71}) with 
singular order four from~(\ref{eq:c72}). Actually, we should consider the normalization  of Eq.~(\ref{eq:c70}) 
with singular order two, but owing to the relation~(\ref{eq:c72}), the two analyses lead to the same conclusion. The 
normalization terms have the form
\begin{gather}
N_{2}(x,y)^{p\sss SE}=:\!A^{\gamma}(x)A^{\rho}(y)\!:\,\big( -i\,N(\d_{x},\d_{y})_{\gamma\rho}\,
\delta^{\sss (4)}(x-y)\big)\,,\nonumber\\
\hat{N}(p)_{\gamma\rho}=\hat{N}(p)^{\sss (0)}_{\gamma\rho}+\hat{N}(p)^{\sss (2)}_{\gamma\rho}+
\hat{N}(p)^{\sss (4)}_{\gamma\rho}\,.
\label{eq:c75}
\end{gather}
Due to the same reasons as pointed out in Sec.~\ref{sec:norm1}, the local normalization can be expressed  in momentum space
through the polynomials
\begin{equation}
\begin{split}
\hat{N}(p)^{\sss (0)}_{\gamma\rho}&=\Psi\,\big[c_{1}\eta_{\gamma\rho}\big]\,,\\
\hat{N}(p)^{\sss (2)}_{\gamma\rho}&=\Psi\,\big[c_{2}p_{\gamma}p_{\rho} +c_{3}\eta_{\gamma\rho}p^2 \big]\,,\\
\hat{N}(p)^{\sss (4)}_{\gamma\rho}&=\Psi\,\big[c_{4}p^2 p_{\gamma}p_{\rho} +c_{5}\eta_{\gamma\rho}p^4\big]\,.
\label{eq:c76}
\end{split}
\end{equation}
Here, $c_{1},\ldots,c_{5}$ are undetermined real numbers and $\Psi:=\kappa^2 \pi/12(2\pi)^5$. Requiring~(\ref{eq:c74}) 
for $\hat{N}(p)^{\sss (i)}_{\gamma\rho}$, $i=0,2,4$, we reduce the freedom in the choice of the  normalization 
parameters in  the polynomials  to
\begin{gather}
\hat{N}(p)^{\sss (0)}_{\gamma\rho}=0\,,\quad \hat{N}(p)^{\sss (2)}_{\gamma\rho}=c_{2}\,\Psi\,
\big[p_{\gamma}p_{\rho}-\eta_{\gamma\rho}\, p^2 \big]\,,\\
\hat{N}(p)^{\sss (4)}_{\gamma\rho}=c_{4}\,\Psi\,\big[p^2 p_{\gamma}p_{\rho}-\eta_{\gamma\rho}p^4\big]\,.
\label{eq:c77}
\end{gather}
The normalized photon self-energy tensor then reads
\begin{gather}
\hat{\Pi}(p)^{\sss N}_{\gamma\rho}=\Big(\frac{p_{\gamma}p_{\rho}}{p^2}-\eta_{\gamma\rho}\Big)\,
\hat{\Pi}(p^2)_{\sss N}\,,\quad\text{with}\nonumber\\
\hat{\Pi}(p^2)_{\sss N}:=\Psi\,\bigg[p^4\,\log\left(\frac{-p^2-i0}{M^2}\right)+c_{2}p^2+c_{4}p^4\bigg] \,.
\label{eq:c78}
\end{gather}
In order to fix the remaining free parameters $c_{2}$ and $c_{4}$, we consider the total photon propagator in momentum
space, defined as the sum of the free photon Feynman propagator and an increasing number of self-energy 
insertions~\cite{scha}
\begin{equation}
\begin{split}
\hat{D}(p)^{\text{tot}}_{\mu\nu}&=+\eta_{\mu\nu}\,\hat{D}_{0}^{\sss F}(p)+
\hat{D}_{0}^{\sss F}(p)\,\tilde{\Pi}(p)^{\sss N}_{\mu\nu}\,\hat{D}_{0}^{\sss F}(p)+\\
&\quad +\hat{D}_{0}^{\sss F}(p)\,\tilde{\Pi}(p)^{{\sss N}\: \lambda}_{\mu}\,\hat{D}_{0}^{\sss F}(p)\,
\tilde{\Pi}(p)^{\sss N}_{\lambda\nu}\,\hat{D}_{0}^{\sss F}(p)+\ldots\\
&=\hat{D}_{0}^{\sss F}(p)\,\big[\eta_{\mu\nu}+\tilde{\Pi}(p)^{{\sss N} \: \lambda}_{\mu}\,
\hat{D}(p)^{\text{tot}}_{\lambda\nu}\big]\,,
\label{eq:c79}
\end{split}
\end{equation}
with $\tilde{\Pi}(p)^{\sss N}_{\mu\nu}:=(2\pi)^4 \hat{\Pi}(p)_{\mu\nu}^{\sss N}$. Multiplying~(\ref{eq:c79}) with
$\hat{D}_{0}^{\sss F}(p)^{-1}$ and with $\eta^{\nu}_{\ \rho}$ we obtain
\begin{equation}
\big[\eta_{\mu}^{\ \lambda}\,\hat{D}_{0}^{\sss F}(p)^{-1}-\tilde{\Pi}(p)^{{\sss N} \: \lambda}_{\mu}\big]\,
\hat{D}(p)^{\text{tot}}_{\lambda\rho}=\eta_{\mu\rho}\,.
\label{eq:c80}
\end{equation}
The inverse of the total photon propagator is then
\begin{equation}
(\hat{D}(p)_{\text{tot}}^{-1})_{\mu\nu}=(2\pi)^2 \,\big[\eta_{\mu\nu}(-p^2 -i0)-(2\pi)^2\,
\hat{\Pi}(p)^{\sss N}_{\mu\nu}\big] \,.
\label{eq:c81}
\end{equation}
Therefore, inverting the expression above, the total propagator reads
\begin{equation}
\hat{D}(p)_{\text{tot}}^{\mu\nu}=\frac{1}{(2\pi)^2}\,\left[\eta^{\mu\nu}\,\frac{1}{-p^2-i0-\hat{\Pi}(p^2)_{\sss N}}
+\frac{p^{\mu}p^{\nu}}{p^2}\,F(p^2)\right]\,.
\label{eq:c82}
\end{equation}
The explicit form of the function $F(p^2)$ is not important, because the last term vanishes between transversal photon
operators. Photon mass- and coupling constant normalization 
\begin{equation}
\hat{\Pi}(p^2)_{\sss N}\Big|_{p^2=0}=0 \quad \text{and}\quad \frac{\hat{\Pi}(p^2)_{\sss N}}{p^2}\Big|_{p^2=0}=0\,,
\label{eq:c83}
\end{equation}
yield $c_{2}=0$. The last parameter $c_{4}$ is non-essential, because it only shifts the mass scale $M$. Therefore, all
local finite ambiguities in~(\ref{eq:c75}) can be reduced to a single unknown normalization parameter. 

This concludes our
discussion of the loop graphs to second order in causal perturbation theory.

\section*{Acknowledgements}

I would like to thank Prof.~G.~Scharf, Adrian M\"uller and Mark Wellmann for discussions and comments regarding these 
topics.

\end{document}